# Optically Detected Magnetic Resonance Imaging

Aharon Blank,[*] Guy Shapiro,[+] Ran Fischer,[+] Paz London,[+] and David Gershoni[+]

[*]*Schulich Faculty of Chemistry, Technion- Israel Institute of Technology, 32000, Haifa, Israel*

[+]*The Physics Department and the Solid State Institute, Technion- Israel Institute of Technology, 32000, Haifa, Israel*

**Abstract**

Optically detected magnetic resonance (ODMR) provides ultrasensitive means to detect and image a small number of electron and nuclear spins, down to the single spin level with nanoscale resolution. Despite the significant recent progress in this field, it has never been combined with the power of pulsed magnetic resonance imaging (MRI) techniques. Here, we demonstrate for the first time how these two methodologies can be integrated using short pulsed magnetic field gradients to spatially-encode the sample. This results in what we denote as an "optically detected magnetic resonance imaging" (ODMRI) technique. It offers the advantage that the image is acquired in parallel from all parts of the sample, with well-defined three-dimensional point-spread function, and without any loss of spectroscopic information. In addition, this approach may be used in the future for parallel but yet spatially-selective efficient addressing and manipulation of the spins in the sample. Such capabilities are of fundamental importance in the field of quantum spin-based devices and sensors.

The selective control and measurement of a small number of electron spins with high spatial resolution is an experimental capability of fundamental importance that is at the basis of many spin-based quantum information devices and sensor technologies. For example, many suggestions and approaches to spin-based quantum computations (QCs – those making use of various quantum phenomena to solve some challenging computation tasks) require such capabilities as a prerequisite for fabricating an actual useful device.[1,2] Furthermore, the manipulation and measurement of electron spins are not only central to many quantum-proposed devices, but can also be used to fabricate spin-based sensors, for example, for high-resolution mapping of magnetic and electric fields with extreme accuracy.[3,4] One possible approach that can potentially answer these requirements makes use of optically detected magnetic resonance (ODMR).

ODMR is a very sensitive technique that can detect a small number of paramagnetic species - even single spins - in specific systems whose optical transitions are coupled to their magnetic levels.[1] In recent years this technique has picked up a significant momentum, due mainly to the many applications and interesting physics revealed through the ODMR of nitrogen vacancy (NV) centers in diamonds.[2] ODMR can be relatively easily employed in the imaging of heterogeneous samples by making use of well-established conventional optical imaging modalities, such as confocal fluorescence microscopy. However, these approaches still have resolution limitations regarding applications aiming at the nanoscale regime.

Furthermore, optical microscopy cannot provide a solution in cases where selective excitation of some of the paramagnetic centers is required, or where some other "dark" spins (i.e., spins that do not produce luminescence signals) in the vicinity of the optical-active center need to be imaged.[5]

The existing approaches to high resolution (nanoscale) imaging and selective addressing of ODMR-active spins make use of scanning probe microscopy techniques, where a sharp magnetic tip that generates very high static gradients (about $10^6$ T/m) is placed in close proximity to the imaged sample surface.[6,7] This allows limiting the volume that meets the resonance condition; thus, by scanning the tip over the sample surface and with the aid of computational deconvolution algorithms, it makes it possible to obtain spatially-resolved information at the sub-nanometer resolution scale of the sample surface. Furthermore, this also provides a means for selective addressing and controlling of the spins in the limited volume defined by the resonance slice of the magnetic tip.[8] However, while the use of a sharp magnetic tip is very effective, it pose some significant problems with respect to the usefulness of the approach to the applications described above, such as resonance slice with a highly distorted geometry, long sequential scanning of the sample, the use of large static gradients that washes out all high resolution spectroscopic information, and it does not allow for efficient parallel selective control and addressing of the spins in the sample.

Here we introduce a new approach that aims at solving these limitations. Our methodology combines pulsed ODMR detection with a pulsed electron spin resonance (ESR) microimaging technique that relies on the well-established principles of medical magnetic resonance imaging (MRI). This approach, which we call optically detected magnetic resonance imaging (ODMRI), can provide high quality 3D images of the sample without loss of the spectroscopic information acquired in a parallel manner using a well-defined point spread function. In principle, it should also allow for parallel selective addressing and control of individual spin populations out of the large spin ensemble of the sample. Following the explanation of the principles of our approach, we provide its first demonstration. Our demonstration includes the imaging of NV centers in a diamond crystal acquired with micron-scale resolution using ODMR detection combined with pulsed magnetic field gradients for spatial encoding.

A schematic description of our experimental setup is provided in Fig. 1a. It combines our "home-made" pulsed ESR imaging system (detailed in ref [9]) with an optical setup for exciting the NV centers in the diamond and detecting their fluorescence signals. We made use of a microimaging probe head operating at room temperature, based on a dielectric ring resonator (fabricated from a single crystal of $TiO_2$) with resonance frequency of 10.6 GHz.[9-11] A type-Ib synthetic diamond, with a [111] face (Element 6), with estimated NV concentration of ~$10^{17}$ spins /$cm^3$ was placed inside the resonator with its surface aligned along the main static field. The diamond was illuminated with a 50-μm multimode optical fiber (Figs. 1b and 1c).

The principles of our approach are as follows: In general, magnetic resonance imaging relies on the use of magnetic field gradients in order to spatially encode the signal coming from the sample, based on the linear relation between resonance frequency and static field (for electron spins, $\hbar \cdot \omega_0 = \mu_B \cdot g \cdot B_0/2$, where $\omega_0$ is the microwave frequency, $\mu_B$ is the Bohr magneton, $g$ is the electron's g-factor, and $B_0$ is the static field's magnitude). Our approach to spatially encode the sample is to employ so-called pulsed field gradients, which is the common approach in nuclear magnetic resonance (NMR) imaging.[12] In order to adapt this approach to the case of ODMR detection of electron spins we made use of the new pulse sequence shown in Fig. 2. Let us analyze this sequence in detail, step by step, so as to provide an insight into its principles of operation.

The first event in the sequence is a short laser pulse with a duration of ~5 μs, which pumps the population of the NV centers' ground state to the $m_s=0$ sub-state (see inset in Fig. 1d).[4] Following this comes a conventional 2-pulse Hahn echo phase encoding imaging sequence, with phase gradients applied between the π/2 and π pulses in a set of different positive and negative amplitudes, thus creating spatial phase encoding in the acquired echo.[12] The effect of these phase encoding gradient pulses and the image acquisition process in a simple case of 3 spin "points" symmetrically located along the X-axis is shown and explained in Fig. 2. (This is similar to the phase effect of the oscillating magnetic field studied in NV-based AC-magnetometers.[13]) Note the required short time scale of the phase gradients - ; on the order of 150-500 ns since it must be shorter than the spin-spin relaxation time, $T_2$ of the imaged spins. This short time scale challenges most experimental capacities.

A unique aspect of ODMR detection is that it can only detect the spin populations of the $m_s$ magnetic sub-states (i.e., the Z-axis magnetization along the externally applied static field), and not the precessing magnetization in the laboratory XY plane (the coherence), as in conventional pulsed magnetic resonance experiments. We circumvent this deficiency by adding a third MW pulse to the echo sequence that rotates the magnetization from the laboratory XY plane to the Z-axis at the time of the echo. This third pulse is followed immediately by an optical laser pulse, during which the fluorescence is detected showing typically the pattern presented by the red curve in Fig. 2c. The change in the fluorescence signal as compared to its steady-state level provides a measure of the spin populations in the $m_s$ sub-states at the time of the echo.[4] It should be noted that the phase gradients' imaging protocols need the full complex information (magnitude and phase) of the spins' magnetization to enable image generation via Fourier processing (Fig. 2).[12] That is, they require measuring the magnitude of the in-phase (X) and out-of phase (Y) components of the magnetization (I and Q) ***perpendicular*** to the axis of the static field (Z). At first sight, it is not clear how this sort of information can be obtained by ODMR detection, because the signal is proportional to the projection of the magnetization ***along*** the laboratory Z direction (having no phase information). In order to provide this information the sequence in Fig. 2 is repeated twice, once with the detection pulse along the X-axis (phase 0) and then along the Y-axis (phase 90°). This first (second) detection X-pulse (Y-pulse) converts the Y (-X) component of the magnetization into the Z component, leaving the one on the X- (Y-) axis unchanged. The combinations of these two measurements provide the full complex data about the magnetization of the spins precessing at the XY plane at the time of the echo. Another important issue with respect to the sequence in Fig. 2 is the use of ± phase cycle on the first MW pulse, which provides a filtered echo signal, clean of other possible signals (like the free induction decay - FID).[14] Additionally, we implemented the so-called CYCLOPS phase cycling scheme[15] on the entire sequence to improve the balance between the in- and out-of-phase complex signal channels (not shown in Fig. 2).

The typical experimental result based on our ODMRI approach is shown in Fig. 3. The 2D image has 200×64 pixels and was obtained by applying only X and Y phase gradients, where X is a vector perpendicular to the diamond's surface (Fig. 3b). The spatial resolution of the image is ~3.5×5 μm, and can be determined in two ways. First, we can use our empirical knowledge of the strength and duration of the pulsed field gradients applied, and then make use of the expression for image resolution[12] provided by: $\Delta x = \frac{1}{(\gamma/2\pi)\int_t G_{max} dt}$, where γ is the electron gyromagnetic ratio. Alternatively, we can count the number of pixels in the image along the diamond piece, which has a known thickness (340 μm), and deduce the pixel resolution using that information. Both methods produce similar results. The ODMRI image clearly shows the expected fluorescence pattern coming from the light entering the diamond crystal and returning to the fiber, as reflected also in the optical microscopy images shown in Fig. 3b. The image shows some vertical artifacts, possibly originating from lack of synchronization of the phase gradients' imaging sequences with the 50-Hz grid frequency. This issue could clearly be improved in the future, possibly by synchronizing with the grid's 50 Hz, or by modulating the light signal at higher frequencies.

The results obtained here can be greatly improved in terms of image resolution as well, as can be deduced from the following arguments: First, we can safely assume that a single spin can be detected with large SNR during a few minutes of measuring time. This is clearly feasible for NV centers in diamonds; their fluorescence signal, when collected with modest efficiency, can reach many thousands of photons per second,[16] and they obviously can provide more than enough signal when using commercially-available sensitive optical detectors. In addition much stronger pulsed field gradients can be obtained (as we demonstrate i with our Q-band micro-imaging setup[9]). Under such conditions, the limiting factor for the image resolution is just the stability of the detected fluorescence signal. More specifically, in order to obtain spatiality-resolved data for a small population of spins out of the entire sample, one must be able to differentiate between small changes in the fluorescence signal that originate from one specific three-dimensional pixel (voxel). In the images we provide here the signal comes from ~300 voxels, and the resulting signal-to-noise-ratio (SNR) is ~50. This means that in the present setup, for a quite reasonable averaging time (~15 min) one can differentiate (with SNR ~1) between ~300×50=15,000 levels of fluorescence signal. In other words, the value of amplitude of the detected ODMR fluorescence signal can be determined within this averaging time within an accuracy of ~$10^{-4}$. Specifically, with the present setup and ~15 min averaging time, it is possible to dissect the full ODMR-originated fluorescence signal from any NV sample to ~100×100 pixels in a 2D image (or ~34×34×10 voxels in a 3D image), provided the field gradients are strong enough. Therefore, if one starts with a much smaller optically-excited volume, for example, on the order of 1 μm, obtained using an appropriate optical setup, then the expected resolution can be well into the nanometer scale, possibly reaching 10 nm and even better (with longer averaging times and/or a more stable laser source).

One of the most significant potential advantages of our approach with the present setup is that it can provide simultaneous spatial addressing and control of spins. This can be achieved by a pulse sequence in which the gradient pulse coincides with the first π/2 pulse. To that end, one can employ the modern-day capability of arbitrary wave form generators[17] for producing a modified MW pulse that can flip specific spins out of the entire sample or control and manipulate them in an arbitrary manner, taking advantage of the large and short pulse gradient employed.

In this letter we present a first experimental demonstration of a new magnetic resonance imaging approach that combines pulsed-field gradients with ODMR detection. This ODMRI method can be further improved to provide high resolution images of ODMR-active spins with high spatial resolution, potentially much better than optical microscopy resolution. The use of pulsed rather than fixed gradients may also support imaging and spectroscopic characterization of "dark" spins that are detected through the ODMR active spins.[6] In addition, this approach facilities the usage of the full spectroscopic arsenal of arbitrary-shaped pulse sequences that can be used to selectively address and control individual spins in a well-ordered spin array in a parallel fashion. This ability is of importance for quantum information processing applications.

**Acknowledgements**

This work was partially supported by grants # 301/13 and # 1524/12 from the Israeli Science Foundation, and by the Russell Berrie Nanotechnology Institute at the Technion.

**Figure captions**

**Figure 1:** The experimental setup and the sample used in the ODMRI experiments. (a) A "home-made" ESR pulsed microimaging system (described in detail in [9-11]) comprised of a pulsed 6-18-GHz microwave bridge and fast pulse gradient drivers is combined with an ODMR optical setup. The sample is positioned at a static magnetic field of 0.27 T, perpendicular to its face, applied by an electromagnet. The optical setup includes a green cw laser source (532 nm) whose light is gated by a computer controlled acoustic optical modulator (AOM). The laser light goes through a multimode optical fiber attached at normal incidence to the sample surface in our imaging probehead inside the electromagnet. The fluorescence emission is collected by the same fiber and goes to a detector, via a long-pass optical filter. (b and c) The imaging probehead is similar to the one described in ref [9-11], based on a compact MW dielectric resonator into which we insert the diamond sample (with the 111 crystal orientation along the direction of the static field). Panel (c) shows a drawing of the main parts of the probe head, before inserting the dielectric resonator into the gradient coil set. (d) Energy levels diagram for an NV center in diamond. The excitation is performed with a 532-nm laser, and the fluorescence is collected for wavelengths longer than 637 nm. In the absence of external magnetic field this spin 1 system is energetically split (by the zero field splitting parameter $D$), into the ms=0 and ms=±1 due to the dipolar interaction. Under a externally applied static field, the levels are split further by the Zeeman interaction (inset).

**Figure 2:** Pulse sequence for ODMRI. (a) The microwave pulses employed – a simple Hahn echo with an additional pulse to transfer the magnetization to the Z-axis to facilitate the ODMR signal. (b) Timing of the transient magnetic field gradients relative to the MW pulses. (c) Optical excitation pulses for pumping and detection of NV population. (d) Spatial magnetic field offset with ($G_1$) and without ($G_0$) field gradient. (e) and (f) circles indicating rotating frames of reference (one-dimensional example) for electron spins located at three different points in space ($x_1$, $x_2$, and $x_3$) without and with applied field gradient, respectively. (g) Oscillating ODMR echo intensity measurements vs. the pulsed-field gradient magnitude enable real space imaging by Fourier transform.

**Figure 3:** (a) ODMRI image of the fluorescence pattern generated by the fiber illumination of the diamond crystal. (b) Optical image of the diamond in the resonator during laser illumination, showing the fluorescence red signal at the center of the crystal.

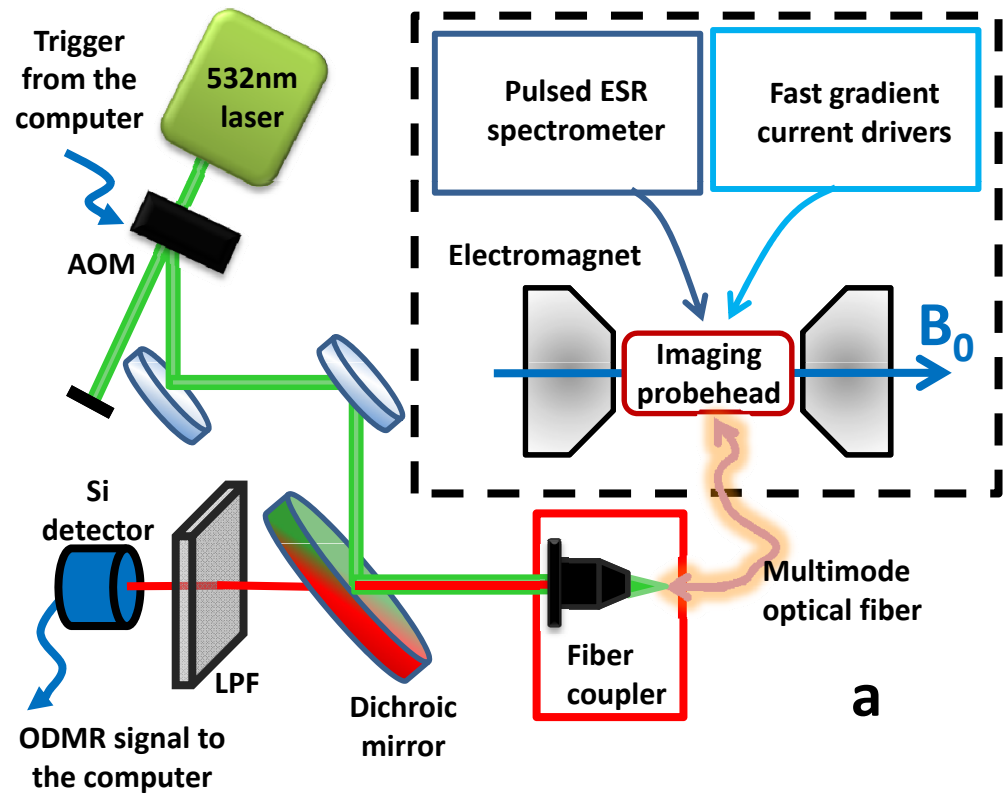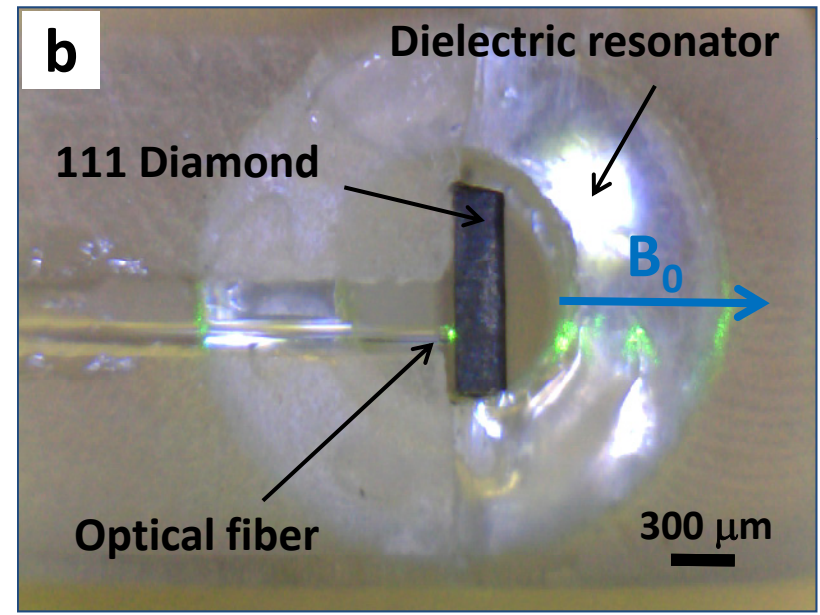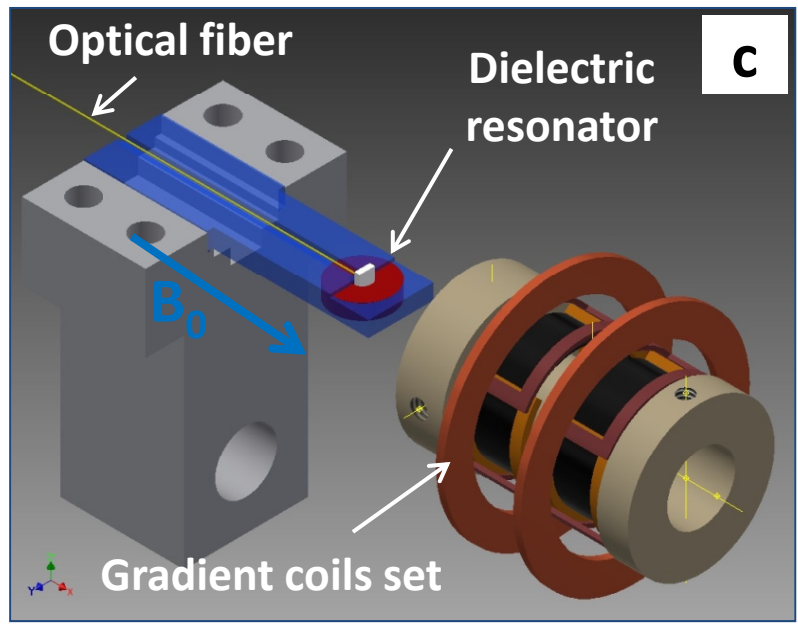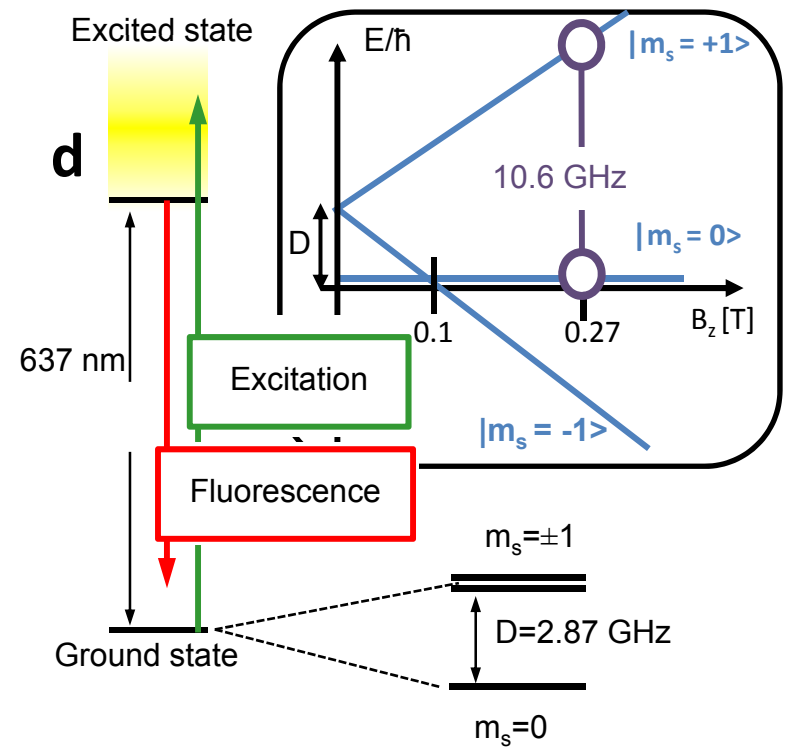

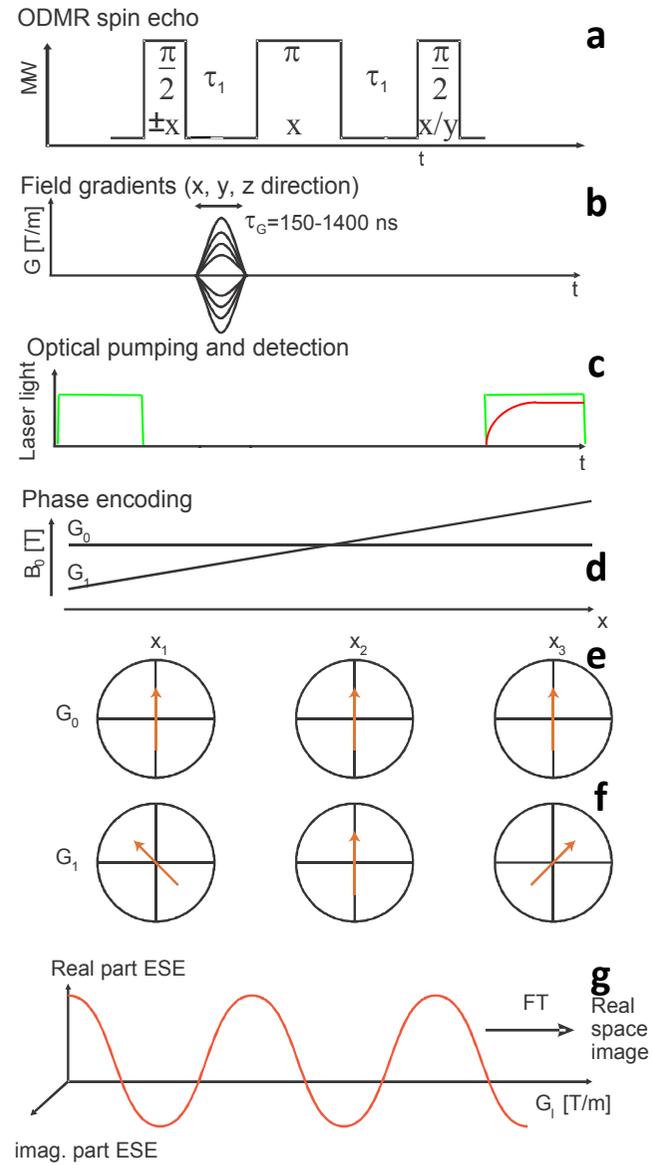

Figure 2

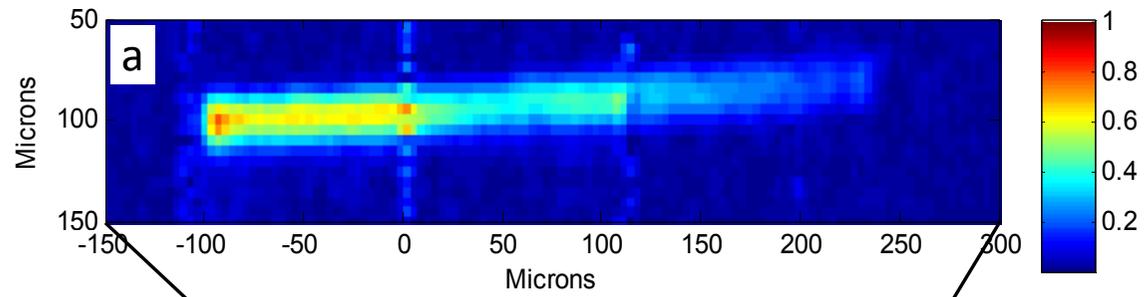
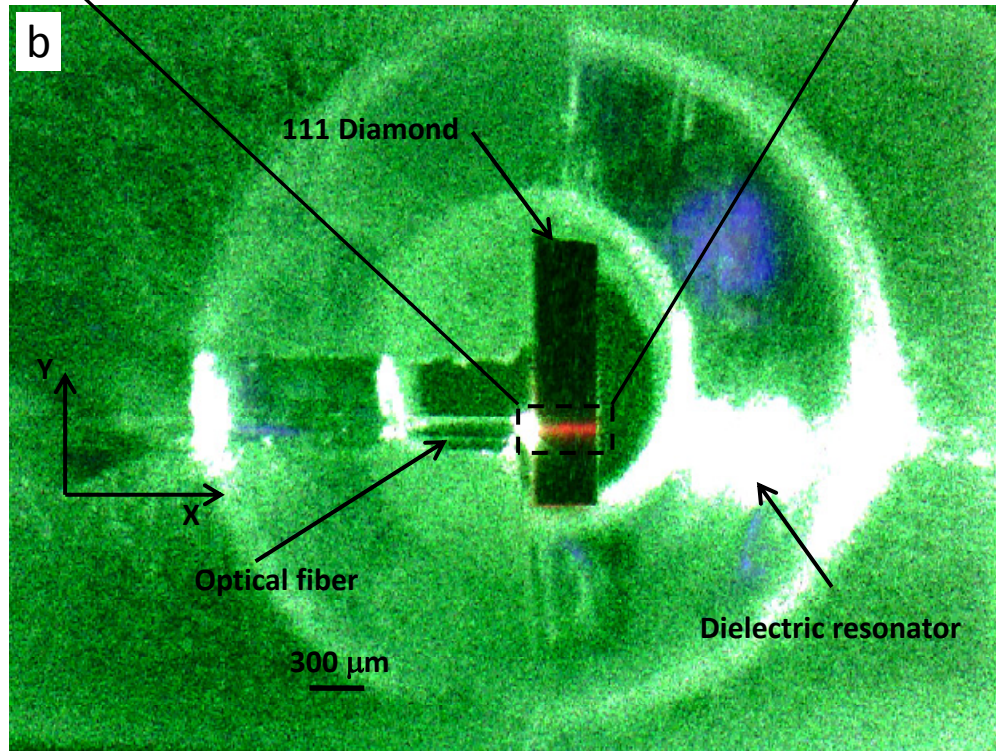

Figure 3